\documentclass[aps,pre,twocolumn,superscriptaddress,showpacs]{revtex4}
\usepackage{graphicx}
\usepackage{verbatim}

\begin{document}

\title{Percolation of randomly distributed growing clusters}
\author{N. Tsakiris} \author{M. Maragakis} \author{K. Kosmidis} \author{P. Argyrakis}
\affiliation{Department of Physics, University of Thessaloniki, 54124 Thessaloniki, Greece}

\date{\today}

\begin{abstract}
We investigate the problem of growing clusters, which is modeled by two dimensional disks and three dimensional droplets. In this model we place a number of seeds on random locations on a lattice with an initial occupation probability, $p$. The seeds simultaneously grow with a constant velocity to form clusters. When two or more clusters eventually touch each other they immediately stop their growth. The probability that such a system will result in a percolating cluster depends on the density of the initially distributed seeds and the dimensionality of the system. For very low initial values of $p$ we find a power law behavior for several properties that we investigate, namely for the size of the largest and second largest cluster, for the probability for a site to belong to the finally formed spanning cluster, and for the mean radius of the finally formed droplets. We report the values of the corresponding scaling exponents. Finally, we show that for very low initial concentration of seeds the final coverage takes a constant value which depends on the system dimensionality.
\end{abstract}
\pacs{64.60.ah, 61.43.Bn, 05.70.Fh, 81.05.Rm}
\maketitle

\section{Introduction}

Percolation theory has drawn a continuous interest from the scientific community for several years \cite{Shante}-\cite{Newman}. It has been studied in a wide variety of systems ranging from lattices \cite{Odor} to complex networks \cite{Newman}. The fields that percolation applies are as diverse as electromagnetism \cite{Bergqvist}-\cite{Hu04}, 
chemistry \cite{Groot}, materials science \cite{Markworth}, geology \cite{McKenzie}, social systems \cite{Palla}, wireless networks \cite{Franceschetti} and many more. In chemistry and materials science it is of major importance for the movement of liquids or gases in porous media. Problems in this area relate to the leakage in seals \cite{Bottiglione} and the gas permeability in cement paste \cite{Galle}.

Various algorithms have been used to simulate the phase transformation kinetics. In many pattern formation models several small spherical seeds are nucleated at a constant rate (homogeneous nucleation). Seeds can also initiate on defects in the case of heterogeneous nucleation. From the simulation point of view the defects are considered as points in the lattice representing the seeds. The seeds once formed are in a metastable phase and grow at a constant velocity as long as there is adequate available material for adsorption. 

Additionally, several models exist that do not allow the adsorption of a new particle in contact with or overlapping with an already adsorbed one. An example is the random sequential adsorption (RSA) model \cite{Evans}. This model has been extensively used for colloid and globular protein adsorption in heterogeneous surfaces. In such systems discrete lattice sites can act as adsorption sites with attractive short range interactions \cite{Adamczykrev}. The jamming coverage and the structure of the particle monolayer as a function of the site coverage and the particle/site size ratio have been studied.

Models studying pattern formation ranging in between these two cases have not been used extensively. Andrienko proposed \cite{Andrienko} the idea of disks and droplets growing at a constant rate on random initial sites over the lattice and stopping once they come in contact. In the so called ``Touch and Stop'' model the droplets grow at a constant rate in all directions (circular in 2D, spherical in 3D). 

The main characteristic of this model is located in the notion that the droplets stop growing after two or more of them come in contact. This can be due to several reasons. In material science it is possible to have a strong surface tension that inhibits the nuclei from taking any shape other than that of a circular or spherical one. Additionally, a significant interacting force between the substrate and the forming droplet can prevent two or more discs from coalescing in the time scale needed for the growth of other islands. 

This problem also relates to the well studied Apollonian packing problem \cite{Kasner} for circles and spheres. In fact it can be considered as a random version of packing with various discrete sizes, where growth velocity is constant but not infinitely large. The Touch and Stop model has also been studied in some variations (random insertion of seeds in time) as a packing limited growth problem \cite{Dodds}. 

\section{Model description}

The system used can be described as follows. Initially, lattices of $10^6$ sites ($1000\times 1000$ for 2D, and $100\times 100\times 100$ for 3D) are randomly populated with seeds of singular size in a non overlapping way. The initial occupation probability of these sites is $p$. At every time step all seeds are investigated once as to the possibility of growing in size instantaneously in all neighboring sites. Investigation sequence is random in order. Each seed is allowed to grow its periphery by one layer (increase the radius by one) provided that there is no overlapping with other growing seeds. Thus each seed becomes an evolving cluster. As soon as two, or more, clusters touch each other, the growth of all of the adjoined clusters stop. The touching disks or droplets can be considered as belonging to the same stable cluster, a cluster that no longer grows over time unless other evolving clusters happen to touch it. Periodic boundary conditions are applied in both the 2D and 3D systems. In order to ensure a smaller statistical error a large number of runs was used (1000 individual configurations).

\begin{figure}
\begin{minipage}[b]{1.0\linewidth}
\centering
\includegraphics[angle=0, width=1.\textwidth]{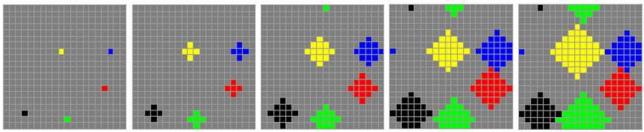}
\caption{Snapshots of the evolution of a 2D system. Each snapshot corresponds to an advance of 1 time unit. The first snapshot is the initial system and the last is the final system configuration, where no percolating cluster is formed. Different colors signify different evolving clusters. At snapshot 4 the evolution of the blue and red droplets stops since they touch and the blue and red droplet now form one stable cluster. All other clusters stop growing at snapshot 5. Periodic boundary conditions are used.}
\label{fig_snaps}
\end{minipage}
\end{figure}

\begin{figure}
\begin{minipage}[b]{1.0\linewidth}
\centering
\includegraphics[angle=0, width=1.\textwidth]{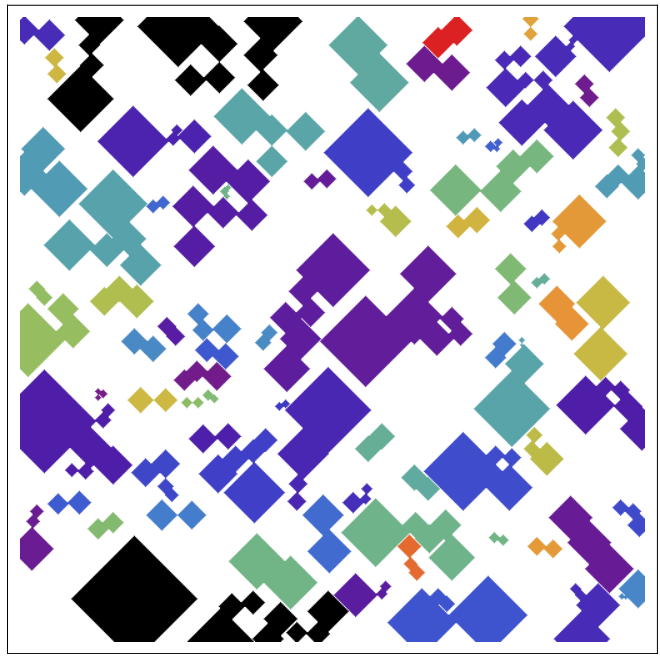}
\caption{Snapshot of the final state of a 2D $500\times500$ system with very low concentration of seeds (p=0.001). Each stable cluster has a different color and the largest is shown in black.}
\label{fig_snaps_larger}
\end{minipage}
\end{figure}

The system continues to evolve until no other cluster can grow in time, so that all of them have at least one adjoining cluster. At this point the final occupation probability of each site is generally larger than $p$ and the system is investigated as to whether it has a spanning cluster using the classic Hoshen-Kopelman algorithm \cite{HoshKop}. The final shape of each disk or droplet of this system is not circular or spherical, in 2D and 3D. It is square and cubic, respectively.  Therefore, it is possible that two evolving clusters have more than just one adjoining sites. In order to have only one adjoining site they must touch at their tips. When two such clusters touch in a part of one of their edges (2D) or facets (3D), the resulting stable cluster formed has two clusters which are connected in more than just one site. Snapshots of a typical finite 2D system evolution can be seen in Fig.~\ref{fig_snaps}, and a larger snapshot of the final state of a system with very low concentration of initial seeds is given in Fig.~\ref{fig_snaps_larger}. It is obvious in Fig.~\ref{fig_snaps_larger} that several clusters are connected in two or more sites.

\section{Discussion}

\begin{figure}
\begin{minipage}[b]{1.0\linewidth}
\centering
\includegraphics[angle=0, width=1.\textwidth]{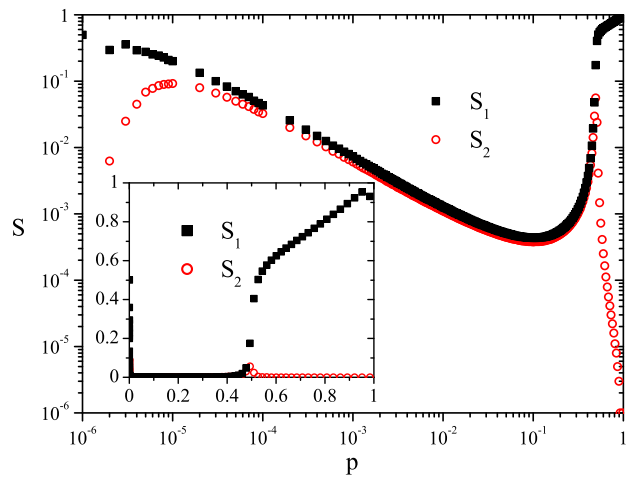}
\caption{Logarithmic representation of the normalized mean largest and second largest cluster sizes, ${S_1}$ (squares) and ${S_2}$ (circles) respectively, over the initial concentration of seeds, $p$, for the 2D case. Inset shows a linear representation of the same where the sharp drop for low initial concentrations and a second transition near $p_c$ are obvious.}
\label{fig_largest2D}
\end{minipage}
\end{figure}

\begin{figure}
\begin{minipage}[b]{1.0\linewidth}
\centering
\includegraphics[angle=0, width=1.\textwidth]{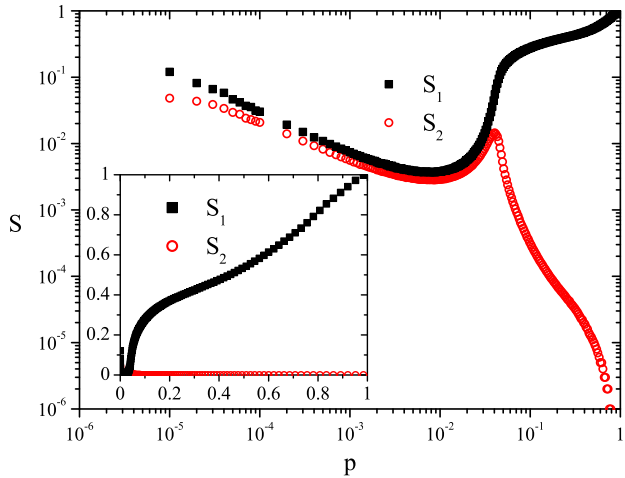}
\caption{Logarithmic representation of the normalized mean largest and second largest cluster sizes, ${S_1}$ (squares) and ${S_2}$ (circles) respectively, over the initial concentration of seeds, $p$, (same as above) for the 3D case. Inset shows a linear representation of the same where the sharp drop for low initial concentrations and a second transition near $p_c$ are obvious.}
\label{fig_largest3D}
\end{minipage}
\end{figure}

Starting from an empty lattice, we vary $p$ in the entire domain $0<p<1$, allow the system to evolve, and we monitor the size of the largest, ${S_1}$, and second largest, ${S_2}$, clusters formed in the final stable configuration (Fig.~\ref{fig_largest2D} and Fig.~\ref{fig_largest3D}). In contrast to the classical percolation model where small concentrations of seeds translates to either isolated sites or very small formed clusters, our model exhibits at first high values for the normalized sizes of the two largest clusters. Initially, the values of ${S_1}$ and ${S_2}$ are quite large, although quickly a sudden drop occurs. This can be explained by the very small number of initial seeds in our system which means that they are randomly placed far apart from one another. This enables them to grow without touching each other for many time steps. Therefore, the largest clusters end up with a high final size value. 

In fact, if we simply have one initial seed, then that one also forms the spanning cluster. This is obviously a finite size effect and is due to the periodic boundary conditions applied. This will cause the growing droplet to eventually touch itself on two opposite vertices of the evolving cluster.

As the initial concentration increases, and until it reaches some system specific value (i.e. $p=0.095\pm 0.003$ for the 2D system), ${S_1}$ and ${S_2}$ reduce. The decrease of the final cluster sizes follows a power law over the initial concentration for a major part of this region ($10^{-5}\leq p\leq5\times10^{-2}$ for the 2D, and $10^{-5}\leq p\leq5\times10^{-3}$ for the 3D) with the exponents having a value of $0.77 \pm 0.02$ and $0.65 \pm 0.02$, respectively. This decrease is caused by the fact that for very small concentrations it is possible to have only several, but large in size, clusters being formed. As the initial concentration is still low but increases, the droplets will now touch each other much faster and the resulting largest clusters will be smaller in size. It is worth mentioning that the application of periodic boundary conditions plays an important role in the calculation of the actual values of the exponents and critical thresholds, especially in the very low and low concentrations of initial seeds. In larger system sizes this effect is reduced.

By further increasing the number of initial seeds and when approaching the half coverage region, many droplets will touch each other at the first few steps, or more commonly be randomly generated at neighboring sites, and therefore not grow at all. The size of the two largest clusters will also start to increase again. As in normal percolation, once the initial concentration reaches a critical value (percolation threshold), $p_c$, all smaller clusters quickly join the one giant cluster formed. At this point the second largest cluster of the system will start to decrease rapidly in size (Fig.~\ref{fig_largest2D} and Fig.~\ref{fig_largest3D}). The initial concentration where this decrease starts is $p=0.497 \pm 0.002$ for the 2D, and $p=0.040 \pm 0.002$ for the 3D case. 
All major clusters from this point on will touch and form one spanning cluster (the giant component) consisting of many adjoined ones. 

\begin{figure}
\begin{minipage}[b]{1.0\linewidth}
\centering
\includegraphics[angle=0, width=1.\textwidth]{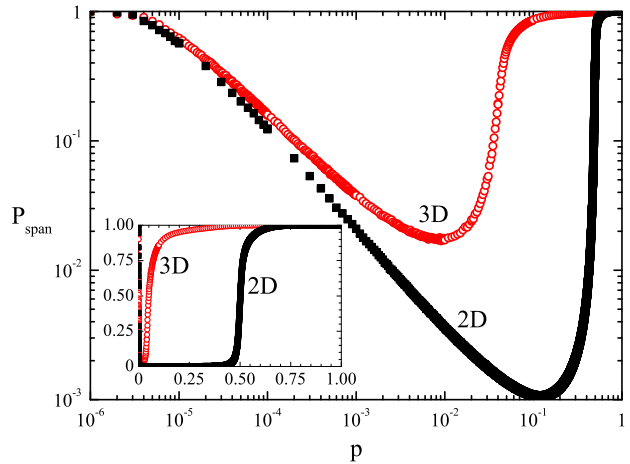}
\caption{Probability of a single occupied site to belong to the spanning cluster over the density of initial seeds shown in a logarithmic representation with the power law area in low values of initial concentration. Squares and circles are for the 2D and 3D case respectively. Inset shows a linear representation where the two phase transitions are clearly shown for the 2D case and 3D cases.}
\label{fig_prob_spanning}
\end{minipage}
\end{figure}

In agreement to the size of the largest clusters, the probability for a spanning cluster to occur, ${P_{span}}$, over the initial concentration exhibits two phase transitions (Fig.~\ref{fig_prob_spanning}). The first ones occur at very low initial concentration, while the second ones occur at a value of ${p\approx0.497}$ and ${p\approx 0.040}$ for the 2D and 3D respectively. Similarly to the largest clusters behavior, ${P_{span}}$ decreases with the increase of the density and reaches a minimum value (for the 2D when $p\approx10^{-3}$ it is $P_{span}\approx10^{-3}$). ${P_{span}}$ also shows a power law behavior for a significant region of low concentration ($10^{-5}\leq p\leq10^{-2}$ for the 2D) with the calculated exponents being approximately $0.77 \pm 0.02$ for the 2D and $0.64 \pm 0.02$ for the 3D. 

\begin{figure}
\begin{minipage}[b]{1.0\linewidth}
\centering
\includegraphics[angle=0, width=1.\textwidth]{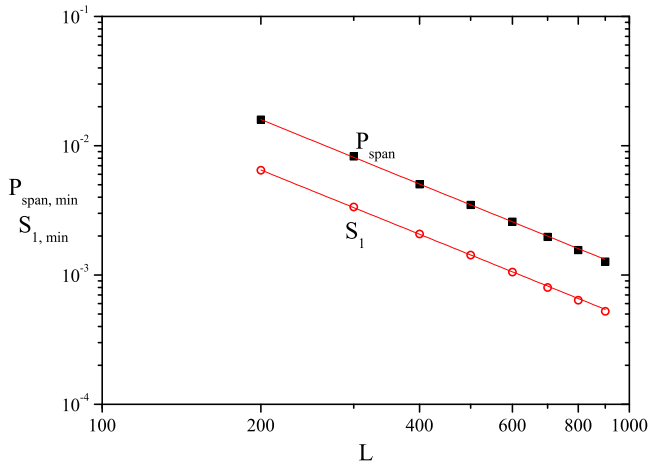}
\caption{Minimum values of ${P_{span}}$ and ${S_1}$ over the size of the system in the 2D case. Squares are values of ${P_{span}}$ and circles are that of ${S_1}$. Lines are best fit.}
\label{fig_min_values}
\end{minipage}
\end{figure}

We performed simulations in various system sizes from $200\times200$ to $1000\times1000$.The shape of the curves observed in Figs.~\ref{fig_largest2D} and~\ref{fig_prob_spanning} is found to be independent of the lattice size with a minimum found on the same value of initial seed concentration. The minimum value of the quantities shown in Figs.~\ref{fig_largest2D} and~\ref{fig_prob_spanning}, exhibits a power law dependence on the size (see Fig.~\ref{fig_min_values}). The exponent of this power is in both cases $1.65 \pm 0.02$.

\begin{figure}
\begin{minipage}[b]{1.0\linewidth}
\centering
\includegraphics[angle=0, width=1.\textwidth]{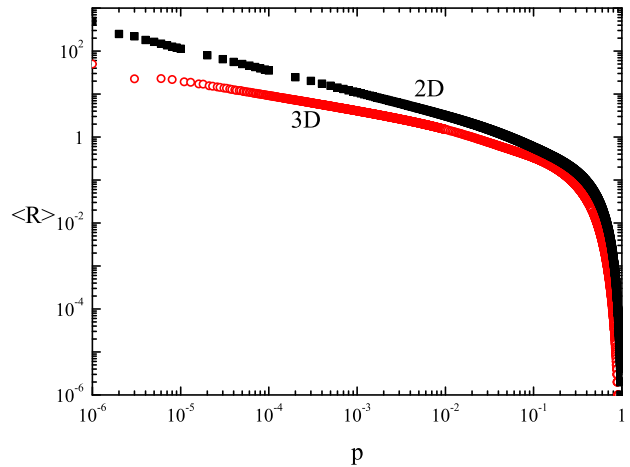}
\caption{Mean radii of the evolving clusters once the system stops. Squares and circles are for the 2D and 3D case respectively. Note that for ${p\geq 0.053}$ for the 2D and ${p\geq 0.020}$ for the 3D case, the value of $\langle R\rangle$ is less than one. This is due to the fact that we measure only the increase of the disks radius and the initial seed is not included.}
\label{fig_aver_size}
\end{minipage}
\end{figure}

As mentioned, cluster evolution stops at various sizes depending on the distance from their neighboring clusters. We calculate the mean radii of the finally formed disks or droplets, $\langle R\rangle$, vs. the initial concentration (Fig.~\ref{fig_aver_size}). We calculate only the increase of the clusters radius and do not include its initial site. For the region of low initial concentration the logarithmic plot reveals a power law behavior of the mean radii which decreases with an exponent $0.52 \pm 0.02$ and $0.37 \pm 0.02$ for the 2D and 3D respectively. 

The exact values of $\langle R\rangle$ for very low concentrations relate to the actual system size dimensions. In the 2D case, the maximum value is $\langle R\rangle=500$ and is found when we have initially only one seed. The corresponding value for our 3D system is $\langle R\rangle=50$. As expected from the application of periodic boundary conditions, it is equal to half the lattices linear dimension. As the concentration increases the mean radius decreases and reaches a value of 1 for ${p\approx0.053}$ and ${p\approx 0.020}$ for the 2D and 3D respectively. For higher concentrations many of the randomly placed initial seeds touch other neighboring clusters at their first time step. Therefore, the neighboring clusters do not grow after this and the average value of the mean radius becomes lower than 1. 

\begin{figure}
\begin{minipage}[b]{1.0\linewidth}
\centering
\includegraphics[angle=0, width=1.\textwidth]{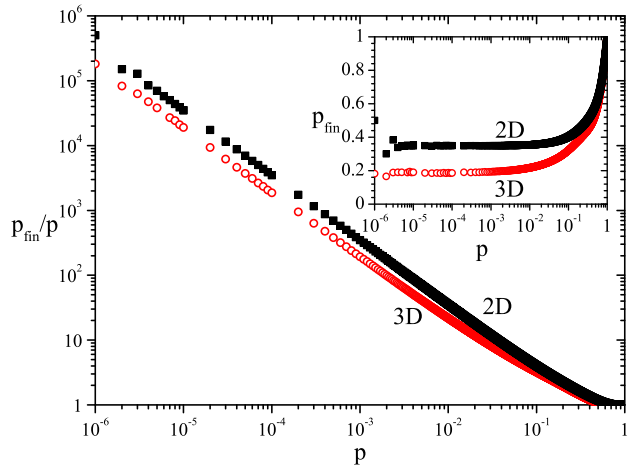}
\caption{The ratio of the final to the initial concentration versus the initial concentration for both the 2D and 3D case. Squares and circles are for the 2D and 3D case respectively. The inset shows that the final concentration remains constant for very low initial concentration for both cases.}
\label{fig_real_occup_2D}
\end{minipage}
\end{figure}

It is also interesting to calculate the ratio of the final lattice coverage to the initial concentration of seeds ${p_{fin}/p}$ in such a system, after all the clusters have stopped expanding. Fig.~\ref{fig_real_occup_2D} shows that the ratio versus the initial concentration exhibits a power-law behavior area in a logarithmic plot where the exponent is calculated to be approximately 1. This means that there is a linear relation between the ratio and the inverse of the initial concentration of seeds. Therefore, we expect a constant final coverage value of the lattice over the initial concentration. Indeed, the inset shows that the final coverage remains constant for both the 2D and the 3D case for very low initial concentration (${p_{fin}=0.35 \pm 0.01}$ for the 2D and ${p_{fin}=0.19 \pm 0.01}$ for the 3D case). These values are found to be independent of the system size.

Overall, the 2D and 3D systems differ qualitatively one with another in the minimum coverage of initial seeds needed for the spanning cluster to occur. In classic percolation, the thresholds of the corresponding 2D and 3D systems studied here are 0.593 and 0.312 respectively. In this model, the 2D thresholds differ much more from the 3D ones, since the 3D threshold is only 0.040, whilst the 2D is 0.497. A 3D system reaches the threshold with a low coverage value. Additionally, the final concentration of the 3D case is much smaller than that of the 2D for very low initial coverages, leaving a large portion of the lattice unoccupied and ending up with a very porous structure.

\section{Conclusion}

We have applied and studied a model of randomly distributed growing clusters from the percolation point of view. The results produced show that there is a region of very low concentration of initially placed seeds where many properties of the system exhibit a power-law behavior. The size of the two largest clusters, the probability to form a spanning cluster and the mean final radii of the evolving clusters all show a power low behavior for approximately 4 orders of magnitude. The minimum values of the size of the largest cluster and the probability to form a spanning cluster also show a power law dependence on the system size. The smaller the system size the higher the probability to have a spanning cluster. Infinite systems should not have spanning clusters formed at low densities. Additionally, we have found that the final coverage in the model for very low initial concentrations is constant and has a value of $0.35$ for the 2D case and $0.19$ for the 3D. 

The accurate determination of the critical percolation threshold of this system requires a more extensive study which is beyond the scope of this manuscript. In \cite{Second_paper}, we calculate the precise values of the percolation threshold and the critical exponents of this transition, and investigate the universality class that this model belongs to. 

This model of randomly distributed growing clusters would be valid for the explanation of properties in problems in material science and technological applications in the cement industry. Variations of this model can also potentially be applied to social networks. We are currently investigating the application of this model to other complex systems. Civilizations can expand until they come in contact with each other at which point they may stop growing for some time because they allocate all available resources to this interaction, i.e. in a war. Such a study would require the application of this model in complex geometries (fractals) and networks. In specific fields, commercial companies can be considered to grow at nearly constant rate until they decide to merge with each other. At such a point their growth is inhibited due to the dedication of large amounts of resources and manpower in the merging process. In both cases the time needed for the growth process can make this circumstantial halting seem as a growth stop when compared to the systems evolutionary speed.

\section{Acknowledgment}

This work was partially supported by the FP6 Project, INTERCONY NMP4-CT-2006-033200.

\newpage \parindent 0 cm \parskip=5mm

\end{document}